\documentclass[aps,rsi,reprint,superscriptaddress,footinbib]{revtex4-1}
\usepackage{hyperref}
\usepackage{graphicx}
\usepackage{amsmath}
\hypersetup{colorlinks=true,linkcolor=blue,citecolor=blue}
\usepackage{xcolor}

\begin{document}

\title{Reduced Thermal Conductivity of Epitaxial GaAs on Si due to Symmetry-breaking Biaxial Strain}
\author{Alejandro Vega-Flick }
\affiliation{Department of Mechanical Engineering, University of California, Santa Barbara, CA 93106, USA}
\author{Daehwan Jung}
\affiliation{Institute for Energy Efficiency, University of California, Santa Barbara, CA 93106, USA}
\affiliation{Center for Opto-Electronics Materials and Devices, Korea Institute of Science and Technology, Seoul 02792, South Korea}
\author{Shengying Yue}
\affiliation{Department of Mechanical Engineering, University of California, Santa Barbara, CA 93106, USA}
\author{John E. Bowers}
\email{bowers@ece.ucsb.edu}
\affiliation{Institute for Energy Efficiency, University of California, Santa Barbara, CA 93106, USA}
\affiliation{Materials Department, University of California, Santa Barbara, CA 93106, USA}
\affiliation{Department of Electrical and Computer Engineering, University of California, Santa Barbara, CA 93106, USA}
\author{Bolin Liao}
\email{bliao@ucsb.edu}
\affiliation{Department of Mechanical Engineering, University of California, Santa Barbara, CA 93106, USA}

\begin{abstract}
Epitaxial growth of III-V semiconductors on Si is a promising route for silicon photonics. Threading dislocations and the residual thermal stress generated during growth are expected to affect the thermal conductivity of the III-V semiconductors, which is crucial for efficient heat dissipation from photonic devices built on this platform. In this work, we combine a non-contact laser-induced transient thermal grating technique with \textit{ab initio} phonon simulations to investigate the in-plane thermal transport of epitaxial GaAs-based buffer layers on Si, employed in the fabrication of III-V quantum dot lasers. 
Surprisingly, we find a significant reduction of the in-plane thermal conductivity of GaAs, up to 19$\%$, as a result of a small in-plane biaxial stress of $\sim$250 MPa. Using \textit{ab initio} phonon calculations, we attribute this effect to the enhancement of phonon-phonon scattering caused by the in-plane biaxial stress, which breaks the cubic crystal symmetry of GaAs. Our results indicate the importance of eliminating the residual thermal stress in the epitaxial III-V layers on Si to avoid the reduction of thermal conductivity and facilitate heat dissipation. Additionally, our results showcase potential means of effectively controlling thermal conductivity of solids with external strain/stress.

\end{abstract}
\maketitle

\section{Introduction}

The field of photonic integrated circuits is rapidly becoming an important contender in the development of optoelectronic devices with improved performance for 
diverse applications such as high-speed telecommunications and information processing \cite{thomson2016roadmap,liu2018photonic}. Among the available integration strategies and material platforms, direct epitaxial growth of III-V compound semiconductors such as GaAs and InP on Si for the fabrication of photonic devices has emerged as a promising direction owing to the reduced cost, the better heat-dissipation capability, the larger available device area and the prospect for scalable manufacturing\cite{liu2018photonic,Sakai1994, wang20111}. However, the epitaxial growth process of III-V materials on Si has its own share of obstacles, mainly caused by the lattice constant mismatch, the formation of anti-phase domains (APD) (due to polarity mismatch\cite{kroemer1980110}) and the thermal expansion mismatch\cite{Sakai1994}. The resulting high threading dislocation density (TDD) and residual thermal stress in the III-V layer are the major factors that negatively impact the efficiency, stability and lifetime of the photonic devices\cite{Sakai1994,liu2015reliability}. Various methods have been developed in order to tackle these problems \cite{liu2018photonic}. Recently, the combination of a thin GaP(45 nm) buffer layer grown on (001) Si substrates (for suppression of APD formation) with buffer structures comprised of GaAs and $\textrm{In}_{x}\textrm{Ga}_{1-x}\textrm{As/GaAs}$ strained superlattices as dislocation filters has led to the fabrication of high efficiency, long lifetime and low threshold III-V quantum dot lasers \cite{jung2018impact, jung2018highly} on Si. In these state-of-the-art devices, the TDD is typically reduced to the level of $\sim 10^6 ~ \textrm{cm}^{-2}$, and the residual thermal stress due to the mismatch in the thermal expansion coefficients of GaAs and Si is $\sim 250$ MPa, corresponding to a strain of roughly 0.16\%\cite{jung2017low, Sugo1989}.

Temperature effects also play an important role in the performance and the lifetime of integrated photonic devices, as an elevated temperature can facilitate the motion and the growth of dislocations, which consequently can lead to device aging and operational malfunction \cite{Sakai1994,thomson2016roadmap}. In this light, efficient heat dissipation from the III-V materials grown on Si is desirable. In principle, both the presence of threading dislocations\cite{zou2002thermal,mion2006accurate} and the residual thermal stress\cite{borca2000thermal,Li2010,Parrish2014,Alam2015} can affect the thermal conductivity of the epitaxial III-V semiconductors grown on Si, directly impacting the thermal management in devices with multilayered structures. Despite a sparse number of previous studies regarding thermal transport in GaAs based devices\cite{Nadri2018}, there has not been direct experimental evaluation of the effect of the TDD and the residual thermal stress on the thermal conductivity of realistic III-V materials grown on Si for photonic integrated circuit applications. 

In this study, we present in-plane thermal transport measurements of 3 $\mu$m thick GaAs based buffer layers employed in the fabrication of III-V quantum dot lasers. The measurements were performed using an optical non-contact, non-destructive method known as laser-induced transient thermal grating (TTG) \cite{Eichler1986, Rogers2000}. We analyzed two multilayered samples with the GaAs based buffer layers and the $\text{In}_{0.1}\text{Ga}_{0.9}\text{As}$/GaAs strained superlattice dislocation filter layers epitaxially grown on different substrates: one on a GaP substrate, and the other on a GaP/Si template (45 nm of GaP on a (001) Si substrate), as shown in Fig.~\ref{figSample}(a). 
Both structures are fundamentally the same; their only difference is the formation of an in-plane residual tensile stress of 250 MPa resulting from the 
growing process (described in section II) on the GaP/Si substrate \cite{jung2018impact}. The stressed buffer layer showed a decrease of 13$\%$ in 
thermal conductivity compared to the unstressed layer. In order to confirm the effect of the residual thermal stress, we further performed TTG measurements on 3 $\mu$m thick GaAs films epitaxially grown on GaAs, GaP and GaP/Si substrates (Fig. \ref{figSample}(b)), and verified a $\sim 19\%$ reduction of the in-plane thermal conductivity. To understand the results, we conducted \textit{ab initio} phonon calculations based on density functional theory (DFT), which predicts a 21$\%$ reduction in the in-plane thermal conductivity of GaAs under a symmetry-breaking 250~MPa biaxial tesile strain, in good agreement with the experimental results.

\section{Sample preparation}
\begin{figure}[h]
\centering
\graphicspath{ {./} }
\includegraphics[width=180pt]{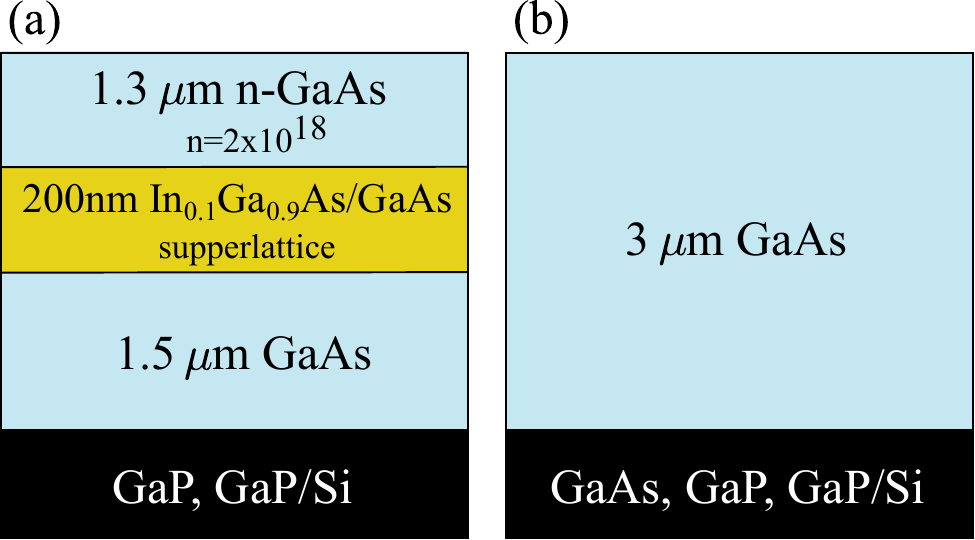}
\caption{Side view schematics of (a) multilayered GaAs based buffer structure grown on GaP, and GaP/Si, and (b) 3 $\mu$m thick films of GaAs grown on three 
different substrates, GaAs, GaP and GaP/Si.}
\label{figSample}
\end{figure}

Detailed description of the growth process can be found elsewhere\cite{jung2017low,Volz2011, Huang2014, Huang2018}, and a brief overview is given here. Two different substrates were selected for growth, GaP, and GaP/Si (see Fig. \ref{figSample}), hereafter referred to as samples s-GaP and s-Si, respectively. The GaP/Si template was provided by $\textrm{NAsP}_{\textrm{III-V}}$ GmbH and consisted of a 775~$\mu$m thick (001) on-axis p-doped Si substrate with a 200~nm thick n-doped Si homo-epitaxial buffer and a subsequent 45~nm thick n-doped GaP nucleation layer deposited by metal-organic chemical vapor deposition\cite{Volz2011}. 
A 1.5~$\mu$m GaAs layer was then grown on both substrates in a solid-source molecular beam epitaxy (MBE), as previously reported\cite{Huang2014,jung2017low}. A thermal annealing cycle was employed after the growth to facilitate dislocation annihilation \cite{Huang2014,jung2017low}. Following this step, a 200~nm $\textrm{In}_{0.1}\textrm{Ga}_{0.9}\textrm{As}$/GaAs strained superlattice layer was grown. This layer is used as dislocation filters for successive film growths \cite{Volz2011, Huang2014, Huang2018}. Finally, 1.3~$\mu$m of GaAs (doped $n \sim 2\times 10^{18}~\textrm{cm}^{-3}$) was grown, providing a template for further III-V device fabrication. The TDD of $7\times 10^{7}$~cm$^{-2}$ and $6\times 10^{7}$~cm$^{-2}$ were measured for the top GaAs buffer layer in samples s-GaP and s-Si, respectively, using electron channeling contrast imaging (ECCI) technique\cite{jung2018impact}. There is an additional in-plane biaxial residual thermal stress of 250 MPa in sample s-Si due to the mismatch of the thermal expansion coefficients of GaAs and Si. This residual stress is absent in sample s-GaP because of the matching thermal expansion coefficients of GaAs and GaP. The residual thermal stress was determined by measuring the red shift of the photoluminescence peak of the GaAs layers\cite{jung2017low, Sugo1989}. In addition to samples s-GaP and s-Si, a set of three GaAs films of 3 $\mu$m thickness were grown on GaAs, GaP and GaP/Si (see Fig. \ref{figSample}(b)) substrates using MBE under the same growth conditions as s-GaP and s-Si. The GaAs film grown on GaP/Si also shows the in-plane residual tensile stress of 250 MPa.

\section{Thermal transport measurements and calculations}

\subsection{Experimental methodology}

\begin{figure*}
\centering
\graphicspath{ {./} }
\includegraphics[width=330pt]{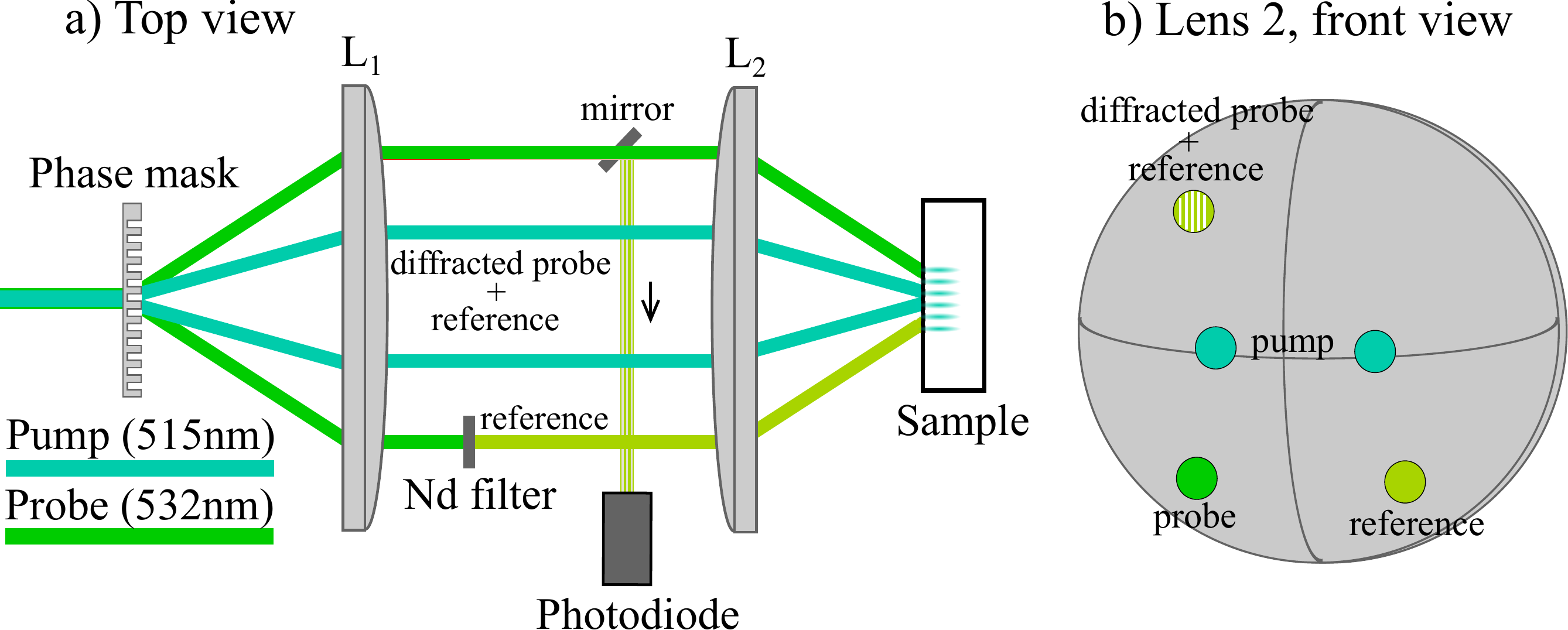}
\caption{Schematic of the TTG setup with optical heterodyne detection. An optical diffraction grating (phase mask) separates the pump and 
probe beams into $\pm$1 orders. One arm of the probe beam is attenuated with a neutral density filter to serve as a local oscillator 
(reference) for heterodyne detection. The pair of probe/reference beams are recombined at the sample and directed into a photodiode detector.}
\label{figTG}
\end{figure*}

In-plane thermal transport was measured using the laser-induced TTG technique. Figure \ref{figTG} shows a schematic of our TTG setup that includes a heterodyne detection scheme. For extensive details regarding heterodyne in a TTG experiment, we refer the 
readers to the references \cite{Maznev1998, Vega-Flick2015}. Briefly, a transmission optical diffraction grating (also known as a phase mask) is used to split 
the excitation and the probe beams into two pairs. A two lens confocal imaging system is used to recombine the excitation and probe beams onto the sample (the focal lengths were $L_{1}=7.5$~cm and $L_{2}=8.0$~cm, respectively). The excitation pulses (pump beams) are from a femtosecond Yb-doped fiber laser at 1030~nm (Clark-MXR IMPULSE), have 260~fs 
pulse width, 250~kHz repetition rate and are frequency doubled to 515 nm wavelength. The spot diameter at the sample is 100~$\mu$m with a $\sim12$~nJ 
pulse energy. The probe beam is a CW laser with a 532~nm wavelength, 90 $\mu$m spot diameter and $\sim30$~mW power. The two excitation laser pulses are crossed at an angle 2$\theta$ in order to produce an intensity pattern with a periodicity $L_{\textrm{TTG}}=\frac{\lambda}{2\sin{\theta}}$, where $\lambda$ is the optical wavelength. In the case of optically opaque samples, absorption of the laser light creates a spatially periodic temperature profile at the surface, which will remain until the 
thermal energy is redistributed from peak to null. The time dependence of the temperature profile can be monitored by diffracting a probe CW 
laser off of the heated region. One of the probe beams is attenuated and used as the local oscillator (reference). Overlapping the reference and the 
diffracted probe light leads to amplification and linearization of the observed signal (heterodyne detection) \cite{Goodno1998, Maznev1998}, 
and is subsequently monitored using a fast photodiode (Hamamatsu C5658) connected to an oscilloscope (Tektronix TDS784A). The diffraction of the 
probe beam is due to both surface displacement induced by thermal expansion and changes in the reflectivity with respect to periodic temperature profile \cite{Johnson2012, Kading1995}.

By quantitatively analyzing the time dependence of the TTG signal, we can obtain the in-plane thermal diffusivity of the sample, which is the material property that physically determines the speed of heat propagation due to temperature differences and is related to the thermal conductivity $\kappa$ through the expression $\kappa=\rho C D $, where $\rho$ is the density, $C$ is heat capacity and $D$ is the thermal diffusivity. A unique feature of the TTG technique is that the length scale of the spatial heating profile can be conveniently controlled by changing the period of the induced thermal grating, which in turn changes the thermal penetration depth probed by TTG.

The multilayered samples measured in this work were considered as a single film with ``effective'' 
thermal properties, grown on a semi-infinite substrate. In this case, the time evolution of the TTG signal can be modeled by solving both the thermal 
diffusion and thermo-elastic equations with a periodic spatial heating source, as presented in reference \cite{Kading1995}. In the case where in-plane thermal transport is dominated by the film, the solution for the TTG signal simplifies to that of a semi-infinite 
half-space with a thermal diffusivity $D$, and is given by the expression \cite{Johnson2012, Kading1995}

 \begin{equation}
\label{eq1}
\begin{aligned}
I_{\textrm{TTG}}(t) = A \ \textrm{erfc}(q_{\textrm{TTG}}\sqrt{D t})+B,
\end{aligned}
\end{equation}

\noindent where $\textrm{erfc}(x)=(2\pi)^{-1/2}\int_{x}^{\infty}e^{-t^{2}}dt$ is the complementary error function, $q_{\textrm{TTG}}=2\pi/L_{\textrm{TTG}}$ is the TTG wavevector, $L_{\textrm{TTG}}$ is the TTG period, $A$ and $B$ are fitting parameters. Equation \ref{eq1} assumes that the thermoreflectance contribution to the TTG signal is small compared to the surface displacement, which is generally the case for non-metals\cite{Kading1995}.

\begin{figure}[h]
\centering
\graphicspath{ {./} }
\includegraphics[width=240pt]{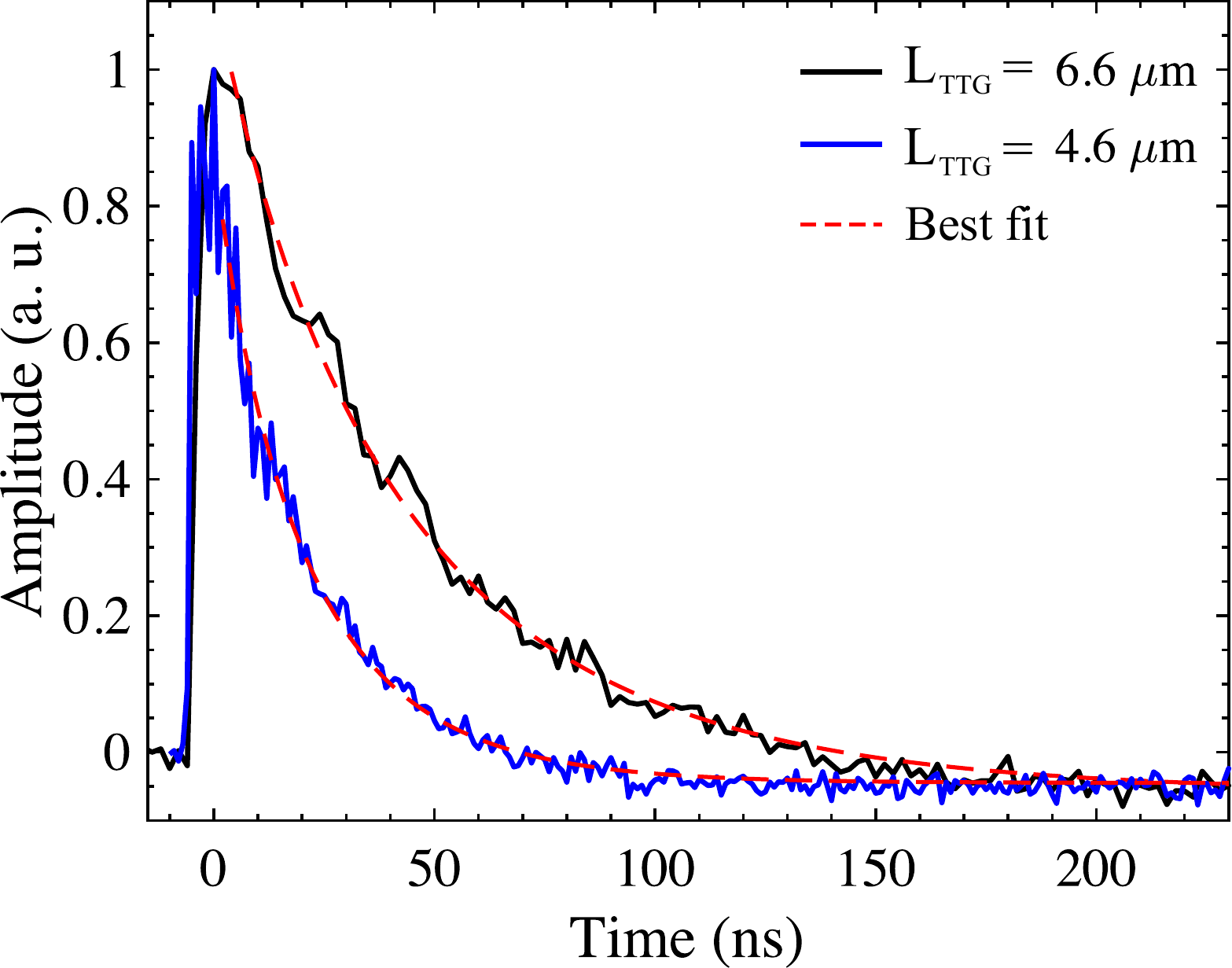}
\caption{Typical TTG time traces obtained for the s-Si sample (GaP/Si substrate) using 6.6 $\mu$m and 4.6 $\mu$m period. The dashed lines correspond to the best fits obtained using Eq. \ref{eq1}.}
\label{figTrace}
\end{figure}

Figure \ref{figTrace} shows typical time traces obtained for sample s-Si using $L_{\textrm{TTG}}$ of 6.6 $\mu$m and 4.6 $\mu$m. 
The dashed lines correspond to the best fits using Eq. \ref{eq1}. As the time scale probed here was tens to hundreds of nanoseconds, the fast dynamics induced by photocarriers, typically happening on the sub-nanosecond time scale, has no effect on the results.

\subsection{Thermal conductivity calculations}

\begin{figure}[h]
\centering
\graphicspath{ {./} }
\includegraphics[width=250pt]{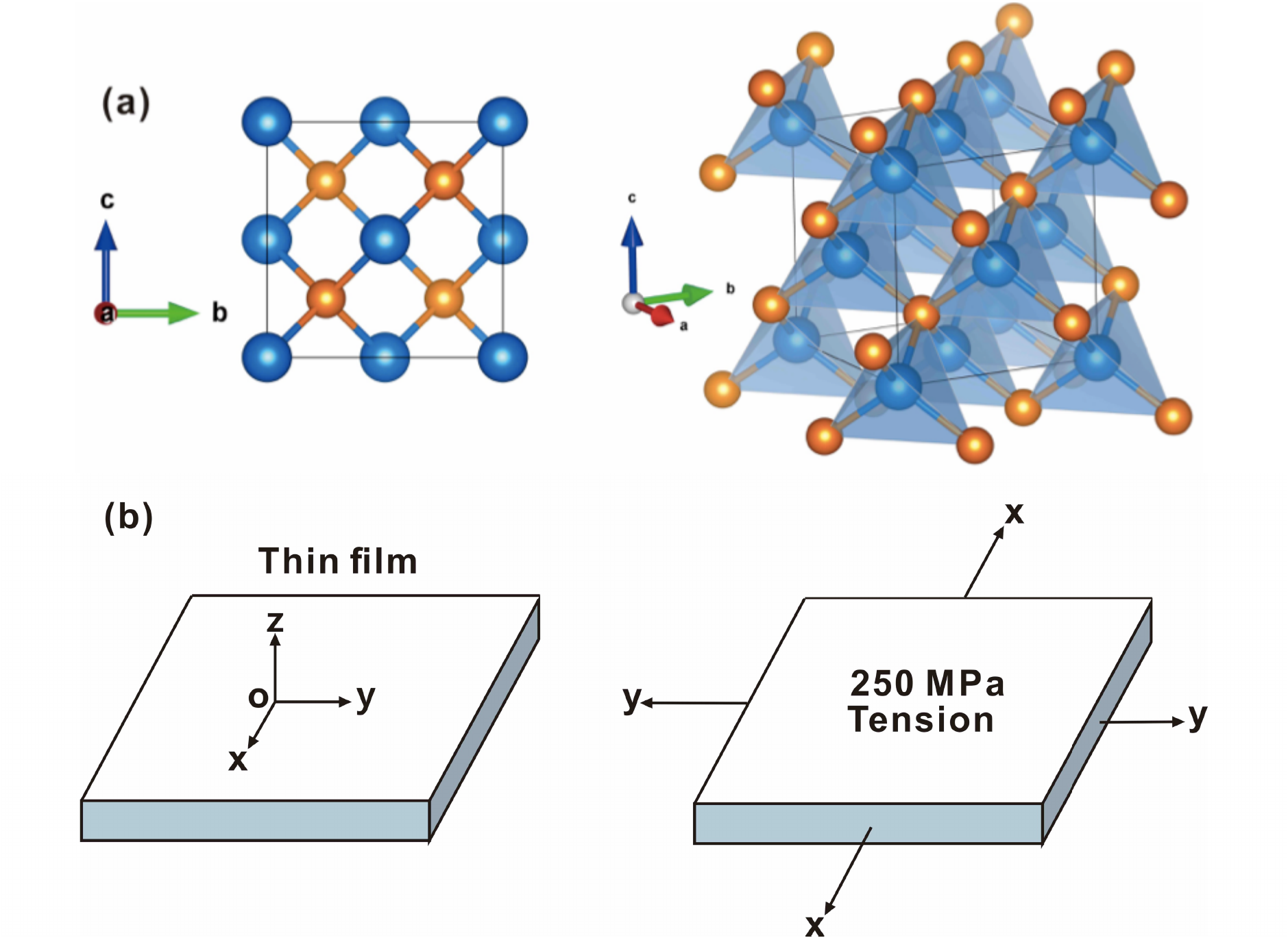}
\caption{(a) Conventional cell for GaAs used in the DFPT calculations and (b) Schematic of the in-plane biaxial tensile stress applied to the films.  }
\label{figCalc1}
\end{figure}

In order to elucidate the effect that the residual thermal stress has on thermal transport, we also performed \textit{ab initio} thermal conductivity calculations of GaAs with or without the in-plane biaxial strain. 
Under the Boltzmann transport equation (BTE) formalism \cite{bte}, the thermal conductivity can be expressed as 

 \begin{equation}
\label{eq2}
\begin{aligned}
\kappa_{\textrm{L}}=\frac{1}{3}\sum_{\mathbf{q}}\sum_{\nu}C_{\mathbf{q}\nu} v_{\mathbf{q}\nu}^2\tau_{\mathbf{q}\nu},
\end{aligned}
\end{equation}

\noindent where $\mathbf{q}$ and $\nu$ are the phonon wavevector and phonon branch, respectively, $C_{\mathbf{q}\nu}$ is the mode-specific heat capacity, $v_{\mathbf{q}\nu}$ is the group velocity, and 
$\tau_{\mathbf{q}\nu}$ is the phonon lifetime. We applied density functional perturbation theory (DFPT) \cite{DFPT} in order to determine the lattice dynamics and consequently calculate 
the thermal conductivity for stressed and un-stressed GaAs. The technical details regarding the \textit{ab initio} calculations are shown in Appendix A. 
Briefly, using the DFPT method we calculated the harmonic second-order interatomic force constants (IFCs), which we employed to determine the phonon dispersion across the whole Brillouin zone (BZ). From here, the group velocity $v_{\mathbf{q}\nu}$ and the heat capacity $C_{\mathbf{q}\nu}$ were calculated as $v_{\mathbf{q}\nu}=\nabla_{\mathbf{q}}\omega_{\mathbf{q}\nu}$ and $C_{\mathbf{q}\nu}=\hbar \omega_{\mathbf{q}\nu}\frac{\partial n_0}{\partial T}$ 
(where $\omega_{\mathbf{q}\nu}$ is the mode specific phonon frequency, $\hbar$ is the reduced Planck's constant, $n_0$ is the Bose-Einstein distribution and $T$ is the temperature). In the following step, we employed the supercell frozen-phonon approach \cite{esfarjani2011heat} in order to 
calculate the third-order (anharmonic) IFCs. In conjunction with the Fermi's golden rule, the anharmonic IFCs were used to calculate the phonon lifetime $\tau_{\mathbf{q}\nu}$. All 
calculations used a conventional cell, which included 8 atoms (see Fig. \ref{figCalc1}(a)).

We calculated the thermal conductivity of GaAs under two different cases of residual stress: i) 0 Pa (unstressed GaAs) and ii)
250 MPa in-plane biaxial stress (X-Y plane, see Fig.~\ref{figCalc1}(b)). As a control, we also calculated the thermal conductivity of GaAs under an isotropic stress of 250 MPa along all three directions. In the calculation, the isotropic stress was implemented by uniformly scaling the conventional cell until the desirable stress was obtained; the biaxial stress was implemented by uniformly adjusting the lattice constants along the X and Y directions, while relaxing the atom positions in the conventional cell and the lattice constant along the Z direction, until the desired in-plane biaxial stress and zero cross-plane stress were achieved. The optimized structure under stress corresponds to a biaxial strain of 0.15\%, in good agreement with experimental measurements\cite{Sakai1994,jung2017low,Sugo1989}.

\section{Results and discussion}

\begin{figure}[h]
\centering
\graphicspath{ {./} }
\includegraphics[width=200pt]{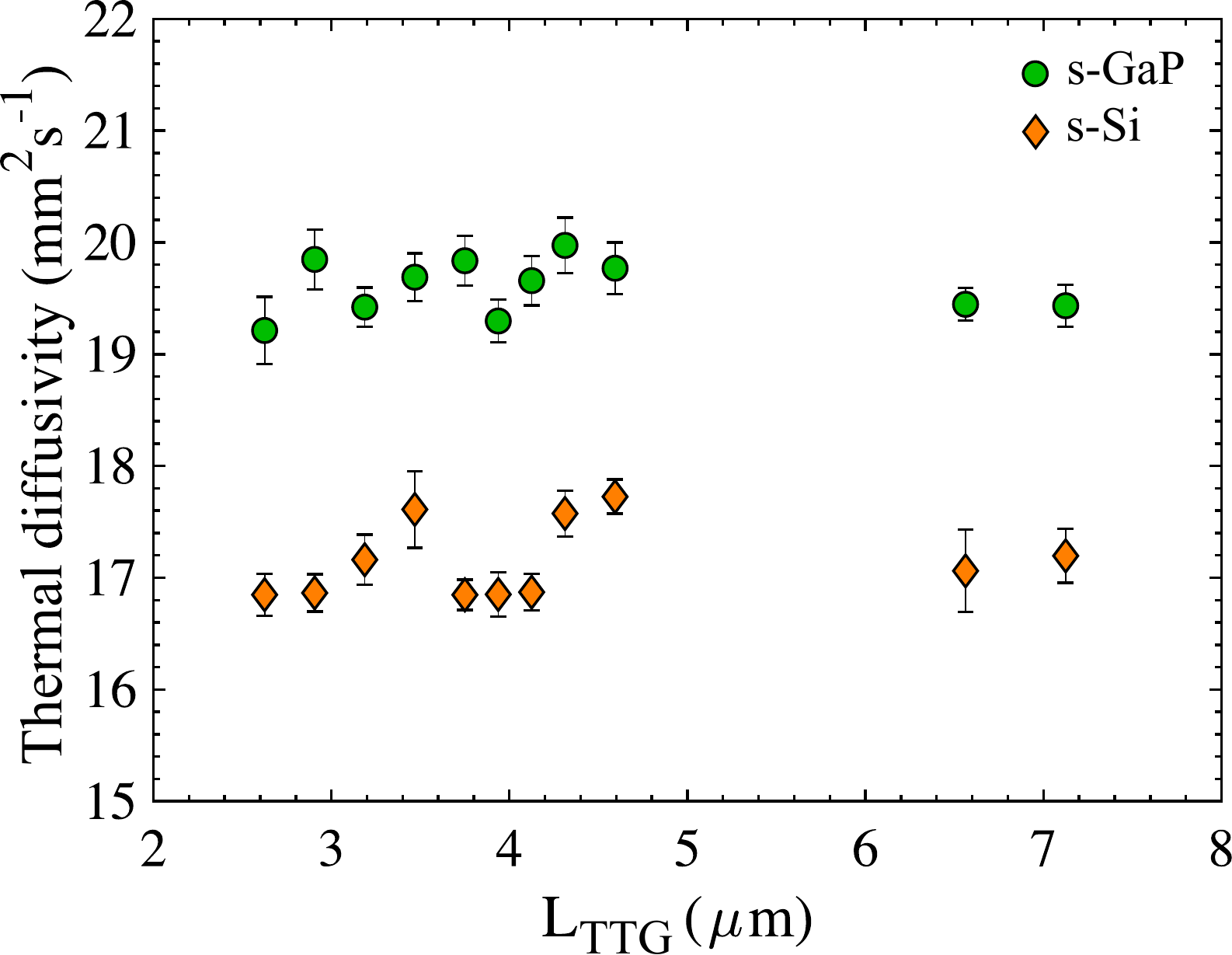}
\caption{Thermal diffusivity values obtained using Eq. \ref{eq1} as a function of the TTG period ($L_{\textrm{TTG}}$) for the multilayer samples s-GaP (circles) and s-Si (squares).}
\label{figThDiff}
\end{figure}

Figure \ref{figThDiff} (a) shows the measured thermal diffusivity values for the s-GaP and s-Si buffer layer samples as a function of the TTG period ($L_{\textrm{TTG}}$) using Eq. \ref{eq1}. The obtained values are independent of $L_{\textrm{TTG}}$, indicating the absence of a substrate effect, i.e. the multilayered structure dominates the in-plane thermal transport, therefore we are effectively measuring the multilayered structures as a bulk semi-infinite material. There is a significant decrease in the in-plane thermal diffusivity of the multilayer structure when it is grown on the GaP/Si substrate ($\sim 13\%$ lower thermal diffusivity). Given the identical structures and similar TDD of the two samples, we attribute the difference in thermal diffusivity to the in-plane residual stress in the sample s-Si. This significant reduction of thermal diffusivity is unexpected given the small magnitude of the stress (0.16\% strain). 

\begin{table}
\centering
\caption{\label{tab:table1} Thermal diffusivity and conductivity values obtained from TTG measurements.}
\small
\begin{tabular}{ccc}
Sample & D(mm$^{2}$/s)&$\kappa$ (W/mK)\\
s-GaP & 19.6$\pm$0.25&-\\
s-Si & 17.1$\pm$0.3&-\\
GaP & 33.3$\pm$1 &54.6$\pm$1.6\\
Si & 59.3$\pm$0.8&105.5$\pm$1.4\\
GaP/Si\footnote{45 nm of GaP grown on Si. The same heat capacity and density of Si was considered for the estimation of $\kappa$.} & 56.7$\pm$1.5&100.9.$\pm$2.7\\
GaAs& 23.5$\pm$0.3&41.3$\pm$0.5\\
GaAs(3$\mu$m)/GaAs & 18.7 $\pm$ 0.62&32.9$\pm$1.1\\
GaAs(3$\mu$m)/GaP & 18.5 $\pm$ 0.63&32.6$\pm$1.1\\
GaAs(3$\mu$m)/GaP(45nm)/Si& 15.2 $\pm$ 0.56&26.7$\pm$1.0\\
\hline
\end{tabular}
\label{table1}
\end{table}

\begin{figure}[h]
\centering
\graphicspath{ {./} }
\includegraphics[width=200pt]{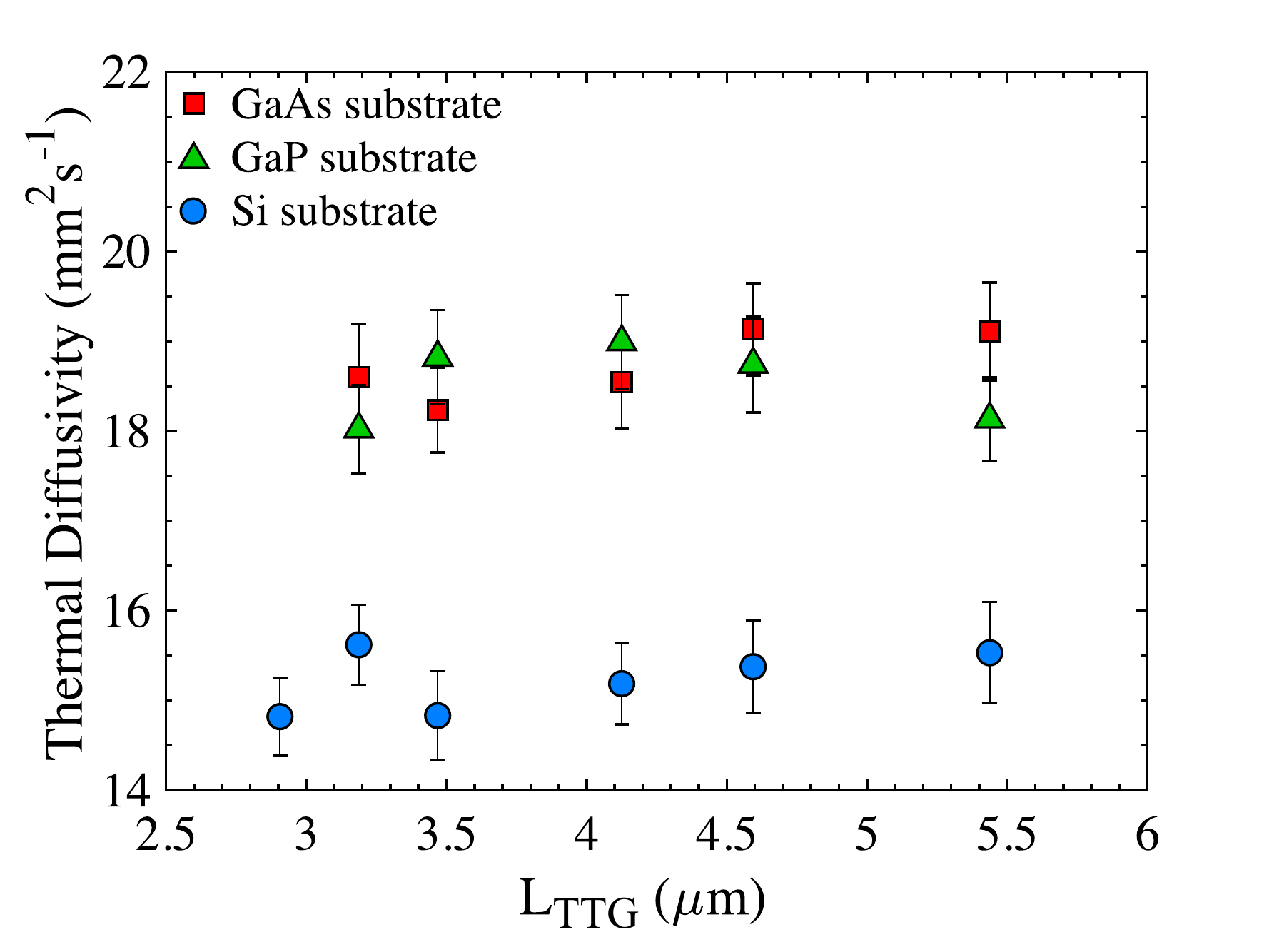}
\caption{Thermal diffusivity values obtained using the complete solution to the thermo-elastic equations 
as a function of the TTG period ($L_{\textrm{TTG}}$) for GaAs (3~$\mu$m) deposited on various substrates: GaAs (squares), 
GaP (triangles) and GaP/Si (45 nm of GaP on a Si substrate, circles).}
\label{figSubs}
\end{figure}

In order to experimentally corroborate the residual stress as the main factor in the reduction of the thermal diffusivity, we measured GaAs films of 3 $\mu$m thickness epitaxially grown on GaAs, GaP and GaP/Si (45 nm of GaP on Si) substrates, as well as the substrates themselves (all obtained values are shown in Table \ref{table1}). The TTG time traces were normalized and analyzed using the complete solution to the thermo-elastic equation \cite{Kading1995} and only the thermal diffusivity of the GaAs film was used as a fitting parameter. All other material properties were taken from literature (see Appendix B). \footnote{The fitting procedure using the complete solution to the thermo-elastic equations was performed with and without the consideration of a thermal boundary resistance between the GaAs film and the substrate. With a resistance value of $R_{Th}=5\times10^{-8}$ mK/W, we found that the fitted thermal diffusivity values were affected by less than 0.5$\%$. The thermal boundary resistance was therefore considered negligible in the rest of the manuscript.} 

Figure \ref{figSubs} shows the obtained thermal diffusivities as a function of $L_{\textrm{TTG}}$. The GaAs film shows similar values for the case of GaP and GaAs substrates ($\sim$ 18.6 mm$^{2}$s$^{-1}$,see Table \ref{table1}). This is expected given that the film is not under residual stress when using GaAs or GaP substrates due to the thermal expansion coefficients of GaAs and GaP being similar \cite{Roesener2013}. Comparing these results to the values obtained using GaAs grown on the GaP/Si substrate (15.2 mm$^{2}$s$^{-1}$), we found a reduction of $\sim$ 19$\%$ in the thermal diffusivity of the stressed film grown on GaP/Si. Additionally, the unstressed film has a lower thermal diffusivity compared to the bulk value ($20\%$ reduction). This can be explained using the Fuchs-Sondheimer theory for thin films, where the effective phonon mean free path (MFP) is reduced due to an increase in the boundary scattering of phonons at the film surfaces \cite{Sondheimer1952,Chen2005}. We also note that the thermal diffusivity of the bare GaAs films (Fig.~\ref{figSample}(b)) without the dislocation filter layers is consistently lower than that of the samples with the dislocation filter layers (Fig.~\ref{figSample}(a)) grown on the same substrates, which can be attributed to the effect of the threading dislocations on phonon transport. It has been known that the threading dislocations can scatter phonons\cite{li2018theory} and reduce the thermal conductivity, for example in GaN\cite{zou2002thermal,mion2006accurate}. A systematic study of the effect of TDD on thermal transport in epitaxial GaAs on Si will be reported in a separate publication.

\begin{figure}[h]
\centering
\graphicspath{ {./} }
\includegraphics[width=210pt]{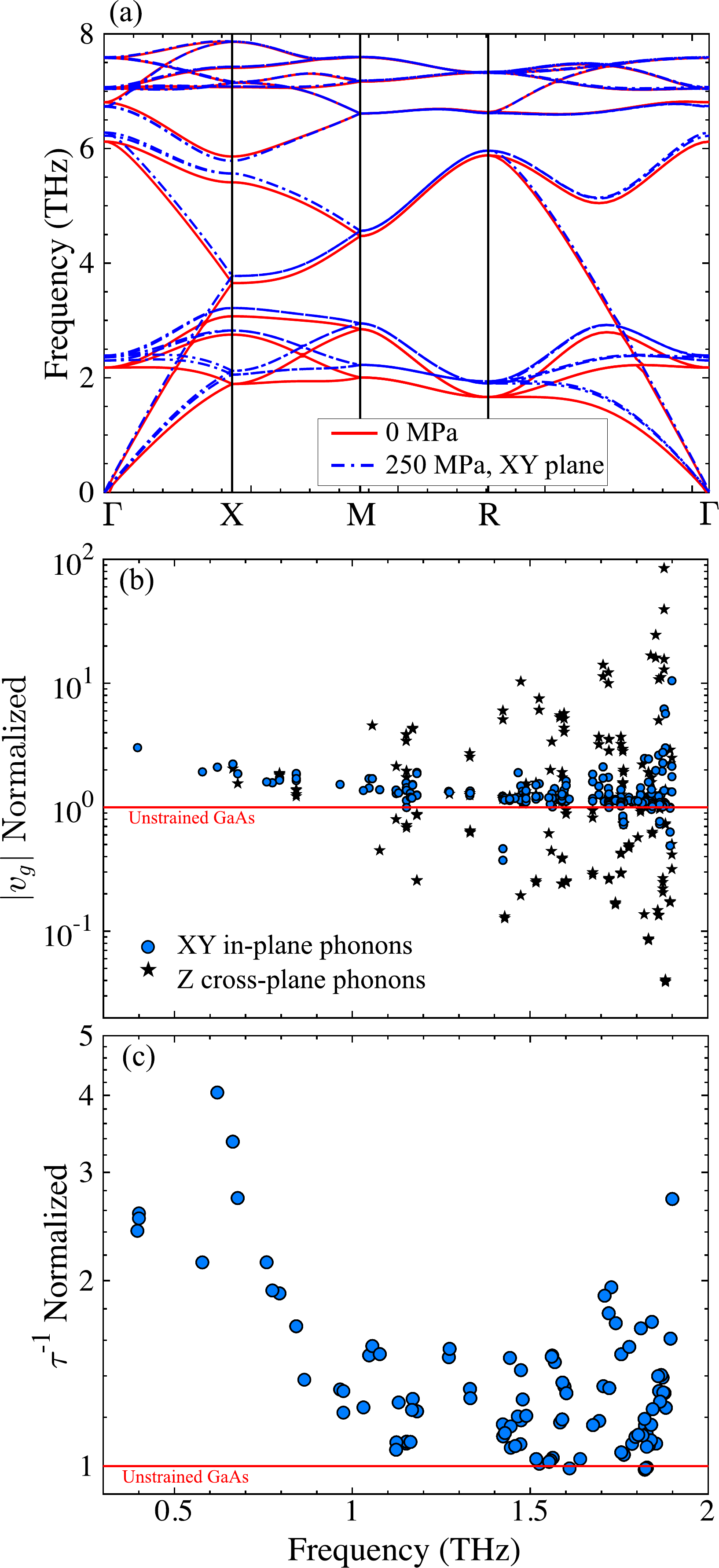}
\caption{(a) Calculated GaAs phonon dispersion across high symmetry directions: solid lines correspond to unstressed GaAs, 
and dashed-dotted lines to GaAs with 250 MPa in-plane biaxial stress. Normalized values as a function of phonon frequency for: (b) group velocity of in-plane and cross-plane phonons, and (c) phonon scattering rate due to phonon-phonon interactions. }
\label{figCalc2}
\end{figure}

\begin{figure}[t]
\centering
\graphicspath{ {./} }
\includegraphics[width=210pt]{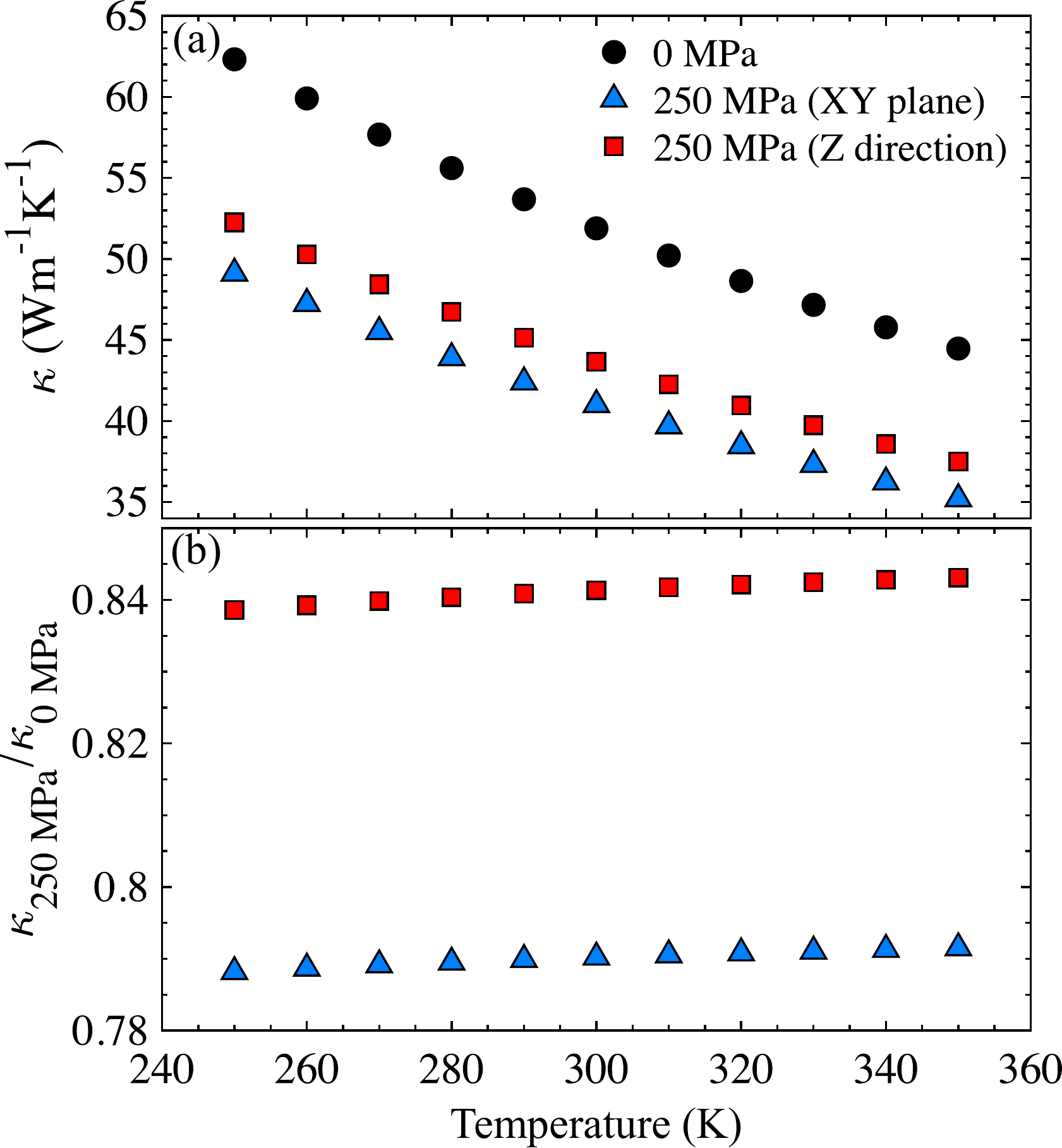}
\caption{(a) Calculated in-plane and cross-plane thermal conductivity of GaAs with and without 250 MPa in-plane biaxial tensile stress as a function of temperature. (b) The in-plane and cross-plane thermal conductivity of stressed GaAs normalized to the thermal conductivity of unstressed GaAs. Circles mark the isotropic thermal conductivity of unstressed GaAs. Triangles and squares mark the in-plane and cross-plane thermal conductivity of stressed GaAs, respectively.}
\label{figCalc3}
\end{figure}

To compare our experimental findings with theory, we performed calculations of the in-plane thermal conductivity of stressed and unstressed GaAs following the procedure described in Section III B. The 250 MPa of tensile stress results in a 0.15\% variation in the lattice constant. This changes the atomic positions in the conventional cell, leading to variations in the phonon band structure. Figure~\ref{figCalc2}(a) shows the phonon dispersion relation comparison between 0 and 250 MPa tensile stress in the XY plane. The high-frequency optical phonon branches show very small changes due to the stress. Only small changes are visible in the $\Gamma-\textrm{X}$ direction. In contrast, the lower frequency acoustic branches show a consistent shift towards higher frequencies. Figure~\ref{figCalc2}(b) and (c) shows normalized values (with respect to the unstressed film) of the heat carrying phonon group velocities and the phonon scattering rates, respectively (phonons with frequencies lower than 2 THz, which are the major heat carriers in GaAs). In the case of the group velocities, cross-plane phonons have a symmetric variation across the base line as a function of frequency, in contrast to in-plane phonons showing a small net increase in the group velocity as a function of frequency, which does not explain the reduced thermal conductivity of GaAs under stress. Strikingly, the small 0.15\% biaxial strain significantly increases the scattering rates of low frequency acoustic phonons, up to a factor of 4, as shown in Fig.~\ref{figCalc2}(c). This is expected to have an important impact on the thermal conductivity, as these low frequency acoustic phonons are the major heat carriers in GaAs. Figure~\ref{figCalc3}(a) shows the calculated isotropic thermal conductivity of unstressed GaAs (circles), the calculated in-plane (triangles) and cross-plane (squares) thermal conductivity of GaAs under the biaxial stress of 250 MPa at different temperatures. The normalized in-plane and cross-plane thermal conductivity of the stressed GaAs with respect to the unstressed GaAs is plotted in Fig.~\ref{figCalc3}(b). The thermal conductivity decreases by an average of $21\%$ and $15.9\%$ for in-plane and cross-plane directions, respectively. The reduction of the in-plane thermal conductivity is caused by the increased phonon scattering rates in the stressed GaAs, and the relative magnitude of the reduction is in good agreement with our experimental results.

The effect of stress/strain on the thermal conductivity of solids has been intensively studied before\cite{picu2003strain,bhowmick2006effect,Li2010,Parrish2014}. The general finding is that tensile stress reduces the thermal conductivity of solids due to the reduction of phonon group velocities and/or specific heat. In previous studies, however, significant reduction of the thermal conductivity typically happens at much higher stress/strain. For example, Parrish et al.\cite{Parrish2014} predicted a 10\% reduction of the thermal conductivity of Si under a tensile strain of 3\%, corresponding to a tensile stress of 7 GPa. Li et al.\cite{Li2010} predicted similar values for bulk Si and diamond. A key difference here is that isotropic strain was applied in these previous studies, whereas in the present study GaAs is under an in-plane biaxial strain. Although isotropic strain modifies the effective ``stiffness'' of the material, the crystal structure of the material is uniformly scaled along all directions and the crystal symmetry is preserved (with the exception of pressure-driven phase transitions\cite{hohensee2015thermal,yue2018electron}). In contrast, in-plane biaxial strain in GaAs also breaks its cubic crystal symmetry with increased lattice constants along the X and Y directions and decreased lattice constant along the Z direction. It is known that high crystal symmetry imposes selection rules on the scattering matrix elements and limits the possible channels of phonon scattering\cite{inui2012group}. In particular, this symmetry-breaking strain effect on electron-phonon scattering in Si and III-V semiconductors has been studied and well understood\cite{Baykan2010, Sjakste2013, Sun2007, Sjakste2006} and the same principle also applies to phonon-phonon scattering. To confirm that the observed significant reduction of thermal conductivity in this work originates from the symmetry-breaking biaxial strain, we also conducted \textit{ab initio} thermal conductivity calculation of GaAs under an isotropic tensile stress of 250 MPa, where the reduction of thermal conductivity was found to be within 2\%. A more rigorous analysis based on group theory is in progress and beyond the scope of this work.

Our findings have multiple implications. On one hand, the significant reduction of the thermal conductivity of epitaxial GaAs on Si due to the residual thermal stress is detrimental to the heat dissipation capability of photonic devices built on this platform. The residual thermal stress is already known to induce motion of the dislocations\cite{Sakai1994,jung2018impact} and reduce the device lifetime, and our new findings provide additional motivation to address the residual thermal stress through rational design of device structures, e.g. by forming high aspect-ratio structures such as micro-ring lasers\cite{jung2018impact}. On the other hand, our results also provide a potential route to design solid-state thermal switches\cite{wehmeyer2017thermal}, whose thermal conductivity can be effectively controlled by external strain/stress.

\section{Conclusions}
In conclusion, we measured the in-plane thermal transport of epitaxial GaAs grown on Si, and discovered a reduction of the thermal diffusivity up to 19\%. By comparing the measurement results of GaAs grown on different substrates, we clarified that the reduction of thermal diffusivity was due to the residual in-plane thermal stress. We further corroborated the result using \textit{ab initio} phonon calculations, and attributed the reduction to enhanced phonon-phonon scattering due to the symmetry-breaking in-plane biaxial stress. Our results reaffirm the importance of addressing the residual thermal stress in epitaxial III-V materials on Si for photonic and electronic applications and may open up new venues towards controlling the thermal conductivity of bulk solids with external means. It will also be of interest to investigate the effect of the TDD and residual thermal stress on the dynamics of hot carriers using time-resolved imaging techniques\cite{liao2017scanning}, as well as dislocation-mediated anisotropic thermal transport\cite{Sun2018}.

\section*{Acknowledgments}
This work is based on research supported by the Academic Senate Faculty Research Grant from University of California, Santa Barbara (UCSB). B. L. acknowledges the support of a Regents' Junior Faculty Fellowship from UCSB.

\section*{Appendix A: Theoretical calculation, technical details}
The \textit{ab initio} calculation was performed using the Vienna ab-initio simulation package (VASP)\cite{vasp01,vasp02} for the DFT and DFPT calculations. For all calculations, we adopted the Perdew-Burke-Ernzerhof (PBE) generalized gradient approximation (GGA)\cite{GGA} as the exchange-correlation functional. 
We employed the pseudopotentials based on the projector augmented wave (PAW)\cite{PAW01,PAW02}. The kinetic energy cutoff of plane-wave functions was set at $\rm 700~eV$ and the tolerance for the energy convergence was $\rm 10^{-8}~eV$. The Monkhorst-Pack\cite{Monkhorst} k-mesh of $\rm 6\times 6\times 6$ was used to sample the Brillouin zone. 
We checked the convergence for the cutoff energy of the plane wave basis and the k-grid density. We used conventional cell which includes 8 atoms in our simulations. 

Details regarding the DFPT calculations of the lattice dynamics are as follows. The harmonic second-order IFC tensors were calculated using the PHONOPY\cite{phonopy}. The non-analytical terms were added to dynamical matrices to capture the polar phonon effects with the Born charges ($\rm Z_{Ga}=2.126$, $\rm Z_{As}=-2.127$) and the dielectric constant ($\rm \epsilon = 12.739$) which were comparable to previous reports\cite{GaAs}. Fine q-grid meshes ($\rm 12 \times 12 \times 12$) were adoped in the DFPT calculations to capture the long-range polar interactions in GaAs. 

The third-order (anharmonic) IFCs were calculated using a supercell frozen-phonon approach. $\rm 2\times 2\times 2$ supercells were used for both calculations with or without the strain. The interatomic interactions were considered up to the 6th nearest neighbours, meaning that the cutoff radius was taken as $\rm \sim 7.05~\AA$. The thermal conductivity, $\kappa_{\textrm{L}}$, was obtained from solving the phonon Boltzmann transport equation\cite{bte} iteratively as implemented in the ShengBTE\cite{ShengBTE} package. 

\section*{Appendix B: Material properties}
Table \ref{table2} shows the literature values for the material properties used to analyze the TTG time traces. The thermal expansion coefficient, Poisson's ratio, shear modulus, 
heat capacity, and density were employed in the calculations of the full solution to the thermo-elastic equations. In the case of the multilayer samples (shown 
in Fig. \ref{figSample}). the TTG data was easily analyzed using Eq. \ref{eq1}, where the only unknown parameter is the effective thermal diffusivity $D$.

\begin{table}[h]
\centering
\caption{\label{tab:table2} Material properties used in the data anaysis.}
\small
\begin{tabular}{cccc}
& GaAs & GaP &Si \\
 $\alpha$ (K$^{-1}$)\footnote{Thermal expansion coefficient}& 5.7$\times$10$^{-6}$&2.6$\times$10$^{-6}$&4.7$\times$10$^{-6}$\\
$\mu$ (GPa)\footnote{Shear modulus}&32.4&62&39.2\\
$\nu$ \footnote{Poisson's ratio}&0.31&0.27&0.31\\
$\rho$ (kgm$^{-3}$)\footnote{Density} & 5320 &2329&4138\\
$C$ (J/K) \footnote{Heat capacity} & 330&704&430\\
\hline
\end{tabular}
\label{table2}
\end{table}

\section*{References}
\bibliography{references.bib}
\end{document}